\documentclass[12pt]{iopart}
\usepackage[utf8]{inputenc}
\usepackage{graphicx}
\usepackage{nicefrac}
\usepackage{iopams}
\usepackage{color}
\usepackage{indentfirst}
\usepackage[usenames,dvipsnames]{xcolor}
\usepackage{hyperref}

\newcommand{\ket}[1]{\left|#1\right>}
\newcommand{\bra}[1]{\left<#1\right|}
\newcommand{\braket}[2] {\left< \left. #1\vphantom{#2} \right| #2 \right>}
\newcommand{\projector}[2] {\left | #1 \right. \rangle \langle \left. #2 \right|}
\renewcommand{\bs}[1]{\boldsymbol{#1}}

\begin{document}

\title{Time evolution of continuous-time quantum walks on dynamical percolation graphs}

\author{Zolt\'an Dar\'azs}
\address{E\"otv\"os University, P\'azm\'any P\'eter s\'et\'any 1/A, H-1117 Budapest, Hungary}
\address{WIGNER RCP, SZFKI, Konkoly-Thege Mikl\'os \'ut 29-33, H-1121 Budapest, Hungary}

\author{Tam\'as Kiss}
\address{WIGNER RCP, SZFKI, Konkoly-Thege Mikl\'os \'ut 29-33, H-1121 Budapest, Hungary}

\begin{abstract}
We study the time evolution of continuous-time quantum walks on randomly changing graphs. At certain moments edges of the graph appear or disappear with a given probability as in percolation. We treat this problem in a strong noise limit. We focus on the case when the time interval between subsequent changes of the graph tends to zero. We derive explicit formulae for the general evolution in this limit. We find that the percolation in this limit causes an effective time rescaling. Independently of the graph and the initial state of the walk, the time is rescaled by the probability of keeping an edge. Both the individual trajectories for a single system and average properties with a superoperator formalism are discussed. We give an analytical proof for our theorem and we also present results from numerical simulations of the phenomena for different graphs. We also analyze the effect of finite step-size on the evolution.
 \end{abstract}


\maketitle

\section{Introduction}
\label{Introduction}

Quantum dynamics, and especially quantum walks on graphs have been a topic of much interest in the past few years \cite{Konno}. The concept of quantum walk was first introduced by Aharonov {\it et al.} \cite{Aharonov} as a discrete-time dynamical system \cite{Meyer}. The continuous-time quantum walk was shown to be an effective tool for quantum computation in the pioneering paper by Farhi and Gutmann \cite{FarhiGut}. Possible applications in quantum information processing \cite{Watrous} triggered thorough research on all aspects of quantum walks \cite{quantumwalks}.

Randomness in graph theory has been an ever developing field since the pioneering works by Erd\H{o}s and R\'enyi \cite{ErdosRenyi}. Today, there is a refined mathematical theory for treating random graphs \cite{Bollobas} and, at the same time, statistical methods borrowed from physics continue to be a motor for research in random network theory \cite{Barabasi}. Applications of classical random graph theory often involve the notion of dynamics on the random graph, describing phenomena ranging from disease spreading \cite{disease} to percolation \cite{percolation}. Random walks provide a simple, fundamental model for dynamics on graphs and random graphs, being one of the first problems in random graph theory and later used throughout its applications \cite{randomwalks}.

Apart from their role as a universal quantum information processing primitive \cite{primitive,Lovett_universal,Childs_exponential,Ambainis_distinctness,Kempe_qrwsearch,Qiang_isomorphism}, continuous-time quantum walks are a conceptually simple, but effective model for transport \cite{Mulken-review}. Randomization of the underlying graph structure is motivated by uncontrolled processes in the physical model, i.e. transport in disordered media \cite{Mulken-random}, similar to percolation in a classical system. Even small changes in the underlying graph can lead to quite dramatic effects for the dynamics of the walk \cite{Xu_extralink}.

In the theory of static percolation, we have a given graph, in which we can keep an edge with probability $\lambda$, or remove it with probability $1-\lambda$. During the time evolution, we keep the graph constant. The ensemble average of a quantity of the system can be calculated as an average over the possible realizations of the percolation graph. In case of continuous-time quantum walks, the static percolation problem has already been considered under the name: statistical networks \cite{statistical-networks, Anishchenko}.

For quantum walks, changing the underlying graph during the time evolution randomly at some rate leads to a different question. In the context of percolation, this is called the dynamical percolation problem \cite{Steifsurvey}. The unitary evolution of the quantum walk is disturbed by the random changes of the graph, leading to a special type of open system causing decoherence, which can be viewed also as a source of noise. Various noise sources leading to interesting effects have been studied in the literature in the context of quantum walks \cite{Chandrashekar-noise, Werner-fluctuations, Werner-random_coin, trapped-lattice}. The first pioneering experiments examining the effect of noise on walks have recently been carried out in photonic experiments \cite{Jex_decoherence, Anderson}. Whether strong noise will eliminate all quantum features is a fundamental question we are interested in. Leung {\it et. al.} \cite{percolation-dtqw} considered first systematically dynamical percolation in quantum walks. They showed that percolation in a discrete time quantum walk even on rather simple graphs can lead to complex time-evolution. On a one dimensional, finite chain it would be hard to calculate the mixing behavior in the longtime limit numerically. For an initially localized walker in a discrete-time quantum walk, applying asymptotic methods, time-dependent oscillations were found in the long time limit, depending on the boundary conditions and the initial internal state \cite{Kollar}. According to a recent paper by Chandrashekar and Busch, on higher dimensional lattices in a directed discrete-time percolated quantum walk transport is critically affected by Anderson localization \cite{Chandrashekar_percolated}.

In this paper we consider the problem of continuous-time quantum walk on a dynamical percolation graph. First, we define the problem of dynamical percolation for continuous-time quantum walks. Unlike in the discrete time case, there is no natural timing for changing the graph structure. One has to introduce a time step for changes. We are interested in the strong noise limit. Strong percolation can happen in two ways: either one chooses a fixed time step and considers the asymptotic limit, or one fixes the evolution time and looks for the continuous percolation limit, when changes in the graph may happen infinitely frequently. For the latter case we will derive analytic expressions leading to rescaling the time and then compare numerically the two limits.

The structure of the paper is the following. In section \ref{sec:CTQW}. we briefly review the basic properties of the continuous-time quantum walk. Then in section \ref{sec:Percolation}. we consider the time evolution of the walk on a percolation graph. We introduce a simple formula for the time evolution and we analytically prove it. We extend our result for ensemble average of quantum walks on percolation graphs with a superoperator formalism in section \ref{sec:Superoperator}. In section \ref{sec:Numerical}. we illustrate our results via some numerical examples. Finally, in section \ref{sec:Finite}. we consider the case when the step-size of the percolation is finite, and we numerically demonstrate that our results can be used for an arbitrary large time if the time-step is small enough.

\section{Continuous-time quantum walk}
\label{sec:CTQW}

Quantum walks have two main types: the discrete- and the continuous-time case. In the discrete-time quantum walk the walker moves on a discrete graph, and changes its position in discrete time intervals.

The concept of continuous-time quantum walk was defined by Farhi and Gutmann \cite{FarhiGut}. We have a $G(V,E)$ undirected graph, and let us denote the vertices with $a=1\dots N$. Then we construct an $N$-dimensional Hilbert space with an orthogonal $\{\ket{a}\}$ basis corresponding to the graph where $a=1\dots N$ and $\braket{a}{b}=\delta_{a,b}$. The unitary time evolution reads
\begin{equation}
	\hat{U} (t)=\e^{-i\hat{H}t}\, ,
\label{eq:unitaryop}
\end{equation}
where the matrix elements of the Hamiltonian are
\begin{eqnarray}
	\langle a |\hat{H}| b \rangle =\left \{ \begin{array}{ll}
	-\gamma, & \textrm{if } |a\rangle \textrm{ and } |b\rangle \textrm{ are neighbouring states,} \\
	0, & \textrm{if } |a\rangle \textrm{ and } |b\rangle \textrm{ not neighbouring,} \\
	k \gamma, & \textrm{if } |a\rangle=|b\rangle.
	\end{array}
\label{eq:hamiltonian}
	\right.
\end{eqnarray}
In the definition of $\hat{H}$ the  $k$ parameter is the degree of the vertex, and $\gamma$ is a time-independent constant. For simplicity, we can choose $\gamma =1$, it only rescales time. The wave function of the system after time $t$ reads 
\begin{equation}
	|\Psi (t)\rangle = \e^{-i\hat{H}t}|\Psi (0)\rangle \, .
\end{equation}

The probability that we measure the walker at the state $\ket{b}$ after time $t$ if we started the walk from the state
$\ket{a}$ can be calculated as
\begin{equation}
    \pi_{b,a}(t)=\left | \bra{b}\e^{-i\hat{H}t}\ket{a} \right | ^2 \, .
\end{equation}

For a classical system, we can describe the time evolution with a very similar formula. The underlying graph is the same, and we can introduce a $\bs{H}$ matrix with a same definition like the Hamiltonian in equation (\ref{eq:hamiltonian}), and with it the probability that the walker goes to the node $b$ from node $a$ after time $t$ reads as the $b,a$ element of the exponential of the $\bs{H}$ matrix, 
\begin{equation}
    p_{b,a}(t) = \left ( \e^{-\bs{H}t} \right )_{b,a} \, .
\label{classp0}
\end{equation}

\section{Quantum walk on a percolation graph}
\label{sec:Percolation}

In the previous sections we briefly introduced the basic model of percolation, and the continuous-time quantum walk. In this section we consider how the dynamics of the continuous-time quantum walk changes if we modify the underlying graph randomly during the walk. We focus on the case when the time interval between the changes tends to zero.

Let us suppose that we have a graph with $N$ edges, and the nodes are connected by a given starting geometry (e.g. ring). Then in given time steps we change the underlying graph randomly. In each step we can keep an edge of the original graph with probability $\lambda$, or discard it with $1-\lambda$. We evolve the system for time $T$ with $\tau$ step-size. In this case the number of the steps is $S=T/\tau$.

For every possible realization of the graph, we can introduce the 
\begin{equation}
    \hat{U}_r(\tau)=\e^{-i\hat{H}_r\tau}
\label{eq:U}
\end{equation}
unitary operator according to (\ref{eq:unitaryop}), where the number of the $r$ realization indexes is $R=2^N$.

Let us suppose that the initial state of the system is $\ket{\psi(0)}$. In each step, we act on the actual state of the system with an $\hat{U}_r(\tau)$ unitary operator. Introduce the following notation: let $\hat{U}_{r_s}(\tau)$ be the time evolution operator according to the randomly generated graph in the $s$th step, in accordance with equation (\ref{eq:hamiltonian}). With this notation, the state at time $T$, after $S$ steps reads 
\begin{eqnarray}
\nonumber
    \ket{\psi(T)}&=&\hat{U}_{r_S}(\tau)\dots \hat{U}_{r_2}(\tau)\hat{U}_{r_1}(\tau) \ket{\psi(0)} \\
	&=& \e^{-i\hat{H}_{r_S}\tau} \dots \e^{-i\hat{H}_{r_2}\tau} \e^{-i\hat{H}_{r_1}\tau} \ket{\psi(0)} \, .
\label{eq:psiTprod}
\end{eqnarray}

Since the $\hat{H}_{r}$ operators do not commute, we can use the Zassenhaus product formula:
\begin{eqnarray}
\nonumber
	\e^{t(A+B)}=\e^{tA}\e^{tB} \prod_{j=2}^q \e^{t^jC^{(A,B)}_j} + \mathcal{O}(t^{q+1}) \, , \\
	\lim_{n\to \infty} \e^{tA_1} \e^{tA_2}\dots  \e^{tA_p}  \e^{t^2C^{\{A_j \}}_2} \dots
	\e^{t^nC^{\{A_j \}}_n}=\e^{t\sum_{j=1}^p A_j} \, ,
\end{eqnarray}
where the $C^{(A,B)}_j$, $C^{\{A_j \}}_n$  parameters are the Zassenhaus exponents \cite{Suzuki, Zassenhaus}. 
We are interested in the $\tau\to 0$ limit, therefore we will neglect the higher order terms in $\tau$:
\begin{equation}
	\e^{-i\hat{H}_{r_p}\tau}\e^{-i\hat{H}_{r_q}\tau} = \e^{-i\tau(\hat{H}_{r_p}+\hat{H}_{r_q})}  +  \mathcal{O}(\tau^2) \, .
\end{equation}
With this approximation, if $\tau$ is small enough, the final state reads
\begin{equation}
       \ket{\psi(T)} = \left ( \e^{-i\tau \sum_{s=1}^S \hat{H}_{r_s}} + \mathcal{O}(\tau) \right )\ket{\psi(0)} \, .
\label{eq:psiT}
\end{equation}

Let us consider the probabilities of the realizations. The probability of the realization that contains $N$ edges is $p^{\{N\}}=\lambda^N$. The probability of a realization, in which one edge is missing is $p^{\{ N-1 \}}=\lambda^{N-1}(1-\lambda)$, and this case is degenerate, because the number of the graphs that contain $N-1$ edges is $N$. It can be seen that the probability of a configuration that contains $k$ edges is $p^{\{k\}}=\lambda^{k}(1-\lambda)^{N-k}$, and the degeneracy is ${N \choose k}$. We can decompose every $\bs{H}_{r}$ matrix in the following way 
\begin{equation}
    \bs{H}_{r}=\sum_{k\in \mathcal{E}_r} \bs{E}_k \, ,
\end{equation}
where $\mathcal{E}_r$ is the set of the $k$ values according to the edges present in the $r$th realization, and $\bs{E}_k$ is a matrix belonging to one edge in the graph. For example, if the $\bs{H}_{r}$ realization contains two edges, the first and the second one, then $\bs{H}_{r}=\bs{E}_1+\bs{E}_2$, and the Hamiltonian of the graph without percolation reads 
\begin{equation}
    \bs{H}=\sum_{k=1}^{N}\bs{E}_k \, .
\end{equation}
Our task is now to evaluate the sum
\begin{equation}
    \bs{S}=\sum_{s=1}^S \bs{H}_{r_s} \, .
\label{eq:sum}
\end{equation}

If $S>R$ then some $\bs{H}_r$ matrix occurs more than once in $\bs{S}$. The probability that we get the $H_r$ matrix a given times from the $S$ steps follows a binomial distribution. If $S\gg R$ then we can use the law of large numbers. We can estimate the number of occurrences of one $\bs{H}_r$ matrix in  the $\bs{S}$ sum as $p_rS + \mathcal{O}(\sqrt{S})$, where $p_r$ is the probability of realizing the graph that belongs to $\bs{H}_r$. The accuracy of our estimation can be calculated from the standard deviation $\sqrt{Sp_r(1-p_r)}$, it leads to an $\mathcal{O}(\sqrt{S})$ error.

If the realization contains $k$ edges, then $p_{r}=p^{\{k\}}$. Now let us calculate, how many times a chosen $\bs{E}_k$ matrix occurs in the $\bs{S}$ sum. This matrix belongs to the $k$ edge in the graph. Choose $j$ edges, where $j=1\dots N-1$, and put them to the graph next to the $k$ edge. We can do it in ${N-1 \choose j}$ ways, therefore there are ${N-1 \choose j}$ graphs that contain $j+1$ edges, and one of these edges is the $k$ edge. The number of occurrences of matrices with $j+1$ edges in the $\bs{S}$ sum from our previous thread is $p^{\{j+1\}} S + \mathcal{O}(\sqrt{S})$. With these results the $\bs{S}$ sum contains the $\bs{E}_k$ matrix
\begin{eqnarray}
\nonumber
    \sum_{j=0}^{N-1}{N-1 \choose j} \left [ p^{\{j+1\}} S + \mathcal{O}(\sqrt{S}) \right ] = \\
	S \sum_{j=0}^{N-1} {N-1 \choose j} \lambda^{j+1} (1-\lambda)^{N-j-1} + 2^{N-1}\mathcal{O}(\sqrt{S}) = 
	S\lambda + \mathcal{O}(\sqrt{S}) 
\end{eqnarray}
times. It is true for every k edges, therefore we can write the $\bs{S}$ sum as
\begin{equation}
	\bs{S}=\sum_{s=1}^S \bs{H}_{r_s}=\left [ S\lambda + \mathcal{O}(\sqrt{S}) \right ] \bs{H} \, .
\end{equation}
This result can be interpreted in the following way: the sum of the $\bs{H}_{r_s}$ matrices can be estimated as $S$ times the $\bs{H}$ matrix for the underlaying graph without percolation rescaled by the probability $\lambda$. Alternatively, one can simply rescale the time instead of scaling the Hamiltonian. We choose the latter, since rescaling the time is convenient when comparing the transition probabilities $\pi_{b,a}(t)$ for the percolated and the percolation free systems.

With the above result, the exponential in equation (\ref{eq:psiT}) can be written as
\begin{equation}
	-i\tau \sum_{s=1}^S \hat{H}_{r_s} = -i T \hat{H} \left (  \lambda + \mathcal{O}(\sqrt{\tau})  \right ) \, .
\end{equation}
Taking the $\tau\to 0$ limit,  we arrive at our main result
\begin{equation}
    \ket{\psi(T)}=\lim_{\tau\to 0} \prod_{k=1}^{\frac{T}{\tau}} \hat{U}_{r_{k}}(\tau)\ket{\psi(0)}=\hat{U}(T\lambda)
     \ket{\psi(0)} \, .
\label{eq:one}
\end{equation}
The effect of a finite $\tau$ step-size will be examined numerically in a later section.

Finally, one can ask what happens if we take the extremal values for the $\lambda$ probability. Both cases are trivial. If $\lambda=0$ then we have a null graph, there is no time evolution, the system stays in the initial state. If $\lambda=1$ then we get the percolation free graph in each step, we get back the percolation free evolution. We  can simply arrive at these results if we write $\lambda=0$ or 1 into the equation (\ref{eq:one}).

\section{Superoperator formalism}
\label{sec:Superoperator}

In this section we would like to consider the average properties of the randomly evolved systems that we introduced in the previous section. The standard way to calculate the ensemble average of a quantity of a system reads 
\begin{equation}
    \langle ... \rangle = \frac{1}{\mathfrak{T}} \sum_{\mathfrak{t}=1}^{\mathfrak{T}} [...]_{\mathfrak{t}} =
    \sum_{\mathfrak{t}=1}^{\mathfrak{T}} p_{\mathfrak{t}} [...]_{\mathfrak{t}} \, ,
\end{equation}
where $\mathfrak{T}$ is the number of the possible trajectories, and $\mathfrak{t}$ denotes one trajectory \cite{Mulken-review}.

In each step, we can keep an edge with probability $\lambda$, or discard it. Let us denote the length of a time interval with $\tau$. Let us consider a graph with $N$ edges, and evolve the system until time $T$. In this case, the graph changes $T/\tau$ times. The graph has $N$ edges, therefore the number of possible graph realizations is $R=2^N$. It means, that for time $T$ the number of the possible trajectories is $\mathfrak{T}=(2^N)^{(T/\tau)}$, because we can choose from $R$ graphs in each step. For a big graph with good time resolution this explicit evaluation becomes very difficult. For example the number of the trajectories for a periodical chain with  $N=15$ nodes with $S=5000$ steps  is over $10^{22577}$ (we use these parameters later for a numerical simulation with the superoperator formalism).

Instead of evaluating all possible trajectories one by one, we can use a superoperator formalism \cite{Novotny, Novotny2, Novotny3}, in which we apply the formerly introduced $\hat{U}_r(\tau)$ unitary operators. In one step, we act on the density operator with the 
\begin{equation}
        \hat{\phi}_{\tau}(\hat{\rho})=\sum_{r=1}^R p_r \hat{U}_r(\tau) \hat{\rho}  \hat{U}^{\dagger}_r(\tau)
\label{eq:superoperator}
\end{equation}
superoperator, where $p_r$ is the probability of the graph realization denoted by $r$ as before, and we sum over all possible realizations. If we evaluate the system for time $T$ with $\tau$ step-size, then we should act $S=T/\tau$ times with the $\hat{\phi}_{\tau}$ superoperator, and the final state reads 
\begin{eqnarray}
	\hat{\rho}(T)= \overbrace{\hat{\phi}_{\tau}( \hat{\phi}_{\tau}( \hat{\phi}_{\tau}(\dots 
	\hat{\phi}_{\tau}}^{S}(\hat{\rho}))) = \phi_{\tau}^{(S)}(\hat{\rho}(0)) \, .
\label{eq:supopseries}
\end{eqnarray}
If we use the (\ref{eq:U}) form of $\hat{U}_r(\tau)$, the final state can be written as
\begin{eqnarray}
\nonumber
    \hat{\rho}(T)&=&\sum_{r_S=1}^R p_{r_S}\hat{U}_{r_S}(\tau) \Bigg ( \sum_{r_{S-1}=1}^R p_{r_{S-1}}
    \hat{U}_{r_{S-1}}(\tau) \bigg ( ... \\
\nonumber
	&&... \sum_{r_1=1}^R p_{r_1} \hat{U}_{r_1}(\tau) \hat{\rho}(0) \hat{U}^{\dag}_{r_1}(\tau) ... \bigg )
	\hat{U}^{\dag}_{r_{S-1}}(\tau)
	\Bigg )\hat{U}^{\dag}_{r_S}(\tau)= \\
	&=& \sum_{r_1... r_S=1}^R p_{r_1}... p_{r_S}\e^{-i\hat{H}_{r_S}\tau}... \e^{-i\hat{H}_{r_1}\tau} \hat{\rho}(0) 
	\e^{i\hat{H}_{r_1}\tau}... \e^{i\hat{H}_{r_S}\tau} \,.
\end{eqnarray}
Here we can use the same treatment in the $\tau\to 0$ limit that we applied in the previous section. In that section we demonstrated, that in the $\tau\to 0$ limit the typical behavior of the possible trajectories is only a time rescaling, therefore if we take the average of the trajectories we again get the time rescaled dynamics. Therefore the act of the superoperator can be written in a very simple form: 
\begin{equation}
    \hat{\rho}(T)=\lim_{\tau\to0}\hat{\phi}_{\tau}^{\left (\frac{T}{\tau}\right )}(\hat{\rho}(0))=\hat{U}(T\lambda)
    \hat{\rho}(0)\hat{U}^{\dagger}(T\lambda) \, .
\label{eq:supopone}
\end{equation}
Formally, a classical system can be treated in a similar way. One can follow the same line of thought and arrive at  the rescaled time. There is, however, a difference between the classical and the quantum systems in the convergence. The quantum systems are very sensitive to the length of the $\tau$ parameter. We discuss its role through examples in the next sections.
	
\section{Numerical examples}
\label{sec:Numerical}

In the previous sections, we considered the dynamics of continuous-time quantum walks on a dynamically percolated graph in a fast changing graph limit. We showed that independently of the underlying graph, in this limit the percolation causes a time rescaling in the dynamics. Our result has been proved both for individual trajectories for a single system and average properties with a superoperator formalism.  In this section we present some numerical examples that demonstrate our results.

In our simulations we calculate the probability that the walker returns to the origin. The return probability is related to the spreading speed and therefore characterizes the transport properties of the walk \cite{Darazs, Zhang, Wan, Agliari}, thus it is a perfect parameter to be used in numerical simulations in order to illustrate our analytical results.

\subsection{Single trajectories on a 2D integer lattice}
\label{sec:trajectory}

\begin{figure}[!ht]
\centering
        \includegraphics[scale=0.9]{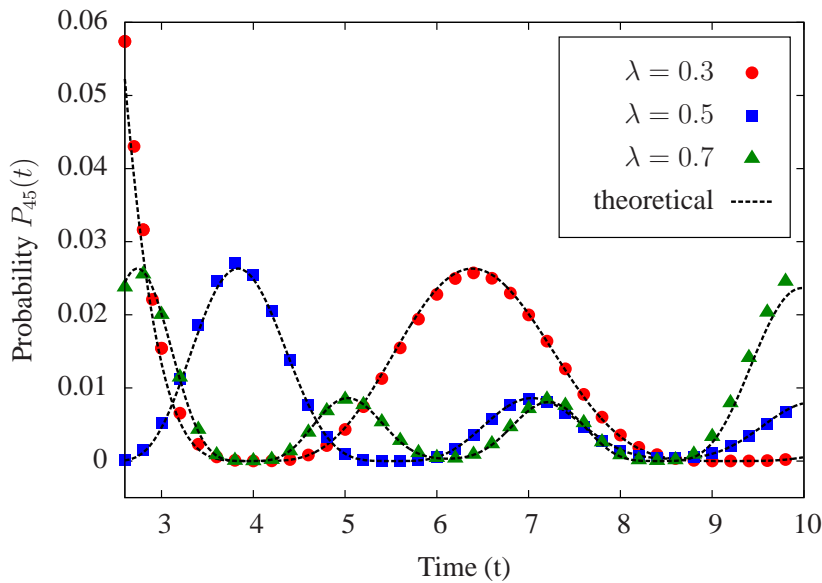}
\caption{The numerical simulation of single trajectories on a $10\times10$ 2D integer lattice in $10^5$ steps with different
	probabilities. The lines show the corresponding analytic estimations (equation (\ref{eq:one})). }
\label{fig:one_system}
\end{figure}

In our first example we consider the continuous-time quantum walk on a percolated 2D integer lattice that contains $N=100$ nodes (figure \ref{fig:one_system}.). We evolve the system with $\tau=10^{-4}$ step-size in $S=10^5$ steps, with a total evolution time $T=10$.

In every step, we randomly choose edges according to the $\lambda$ probability, and we take the $\bs{H}_r$ matrix for this realization. Then we act on the state from the previous step with the unitary operator $\hat{U}_r(\tau)$ according to equation (\ref{eq:U}). With this method we numerically realize the state in equation (\ref{eq:psiTprod}).

We start the system from the $\ket{\Psi(0)}=\ket{45}$ state, it is in the middle of the lattice, and we calculate the probability of being at this node. We use different $\lambda$ values, and we plot both the numerical and the theoretical values according to equation (\ref{eq:one}). The results are shown in figure \ref{fig:one_system}. Incidentally, we note that the shape of the curves are very close to the appropriate Bessel function, which is the analytical solution of the 2D unpercolated lattice \cite{Muelken2D} if we scale them by the $T\to \lambda T$ transformation in the percolation case. This suggests that our main result also holds for infinite 2D integer lattice graphs. 

\subsection{Superoperator formalism on a ring}

\begin{figure}[!ht]
\centering
    \includegraphics[scale=0.9]{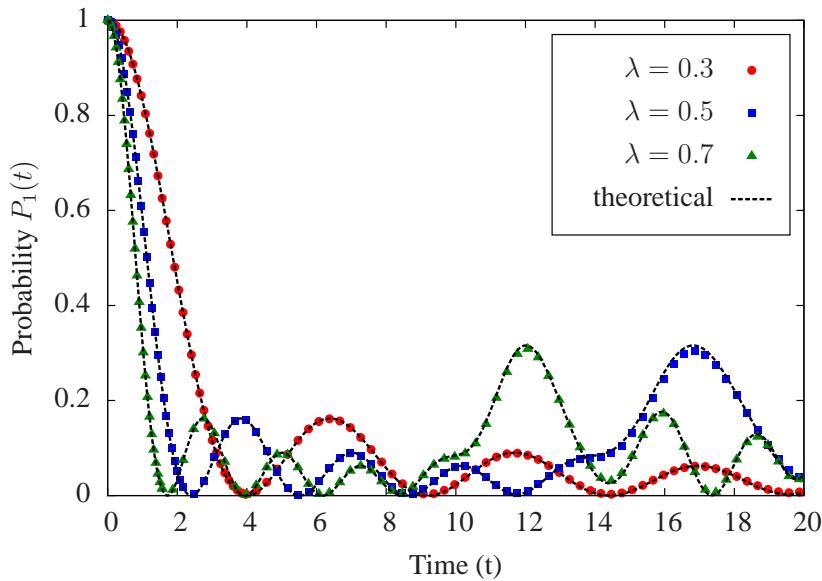}
\caption{Return probability on a periodical chain with $N=15$ nodes numerically calculated by the superoperator formalism. The parameters are $S=5000$ and $\tau=0.004$. }
\label{fig:superoperator}
\end{figure}

In figure \ref{fig:superoperator}. we demonstrate that the equation (\ref{eq:supopone}) gives a very good approximation for the time evolution with finite $\tau$ values if we take an average over the trajectories, i.e. we use the superoperator formalism.

In our simulation, we start with preparing all the possible $\bs{H}_r$ matrices, and then the corresponding $\bs{U}_r(\tau)$ matrices in accordance with equation (\ref{eq:U}). In one step we take the state of the system after the previous step, and we calculate the average over the possible realizations by the definition of the $\hat{\phi}_{\tau}$ superoperator from equation (\ref{eq:superoperator}). This repeated action of the superoperator is the numerical interpretation of equation (\ref{eq:supopseries}). One can explicitly evaluate this algorithm only for relatively small systems, because the number of the matrices scales as $2^N$. By this method we manifest one of the main advantages of the superoperator formalism: we can take an average over the possible realizations in every step, instead of evaluating all the possible trajectories, which is more difficult even for small systems, see section \ref{sec:Superoperator}.

In our simulation we use a periodical chain with $N=15$ nodes, and we evolve the system with $\tau=0.004$ step-size from the $\hat{\rho}(0)=\projector{1}{1}$ initial state in $S=5000$ steps. We use different values for the probability, and we also plot the theoretical line from equation (\ref{eq:supopone}) in figure \ref{fig:superoperator}. Again, we can see that the starting shape of the curves is very close to the $J_0^2(2t)$ Bessel function  \cite{Spacetime}. This fact again suggests that our results are valid for the infinite 1D integer lattice graph.

\subsection{Single trajectory on a random graph}

\begin{figure}[!ht]
\centering
        \includegraphics[scale=0.9]{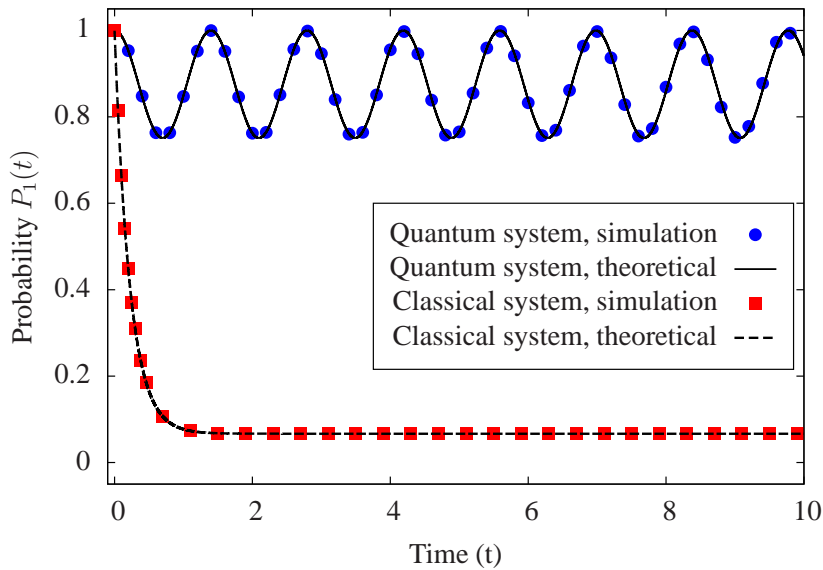}
\caption{Comparison of the return probabilities of the classical and quantum walk on a dynamical percolation full graph. The graph has $N=15$ nodes, and we evolve the system with $\tau=10^{-4}$ step-size in $S=10^5$ steps, and we choose the percolation probability for $\lambda=0.3$. }
\label{fig:complete}
\end{figure}

As a third example, we take a complete graph in figure \ref{fig:complete}. with every possible connection among the $N=15$ nodes. Note that the oscillations in the return probability do not decay after a long time which is a consequence of the nonclassicality of the continuous-time quantum walk.
For our model we calculate one single trajectory, as we describe our method in section \ref{sec:trajectory}. We start the system from the $\hat{\psi}(0)=\ket{1}$ state, and we evolve it with $\tau=10^{-4}$ step-size $S=10^5$ times, and we choose the probability of the percolation for $\lambda=0.3$. We plot our results with (blue) dots.
Our analytical proof is independent of the underlying graph, therefore we can apply it to this graph too. Every connection is enabled in this system, thus the underlying structure is a percolated complete graph. It is a general graph, consequently the analytical return probability can be calculated straightforwardly.
The Hamiltonian of the system can be represented in the $\boldsymbol{H}=N\boldsymbol{E}-\boldsymbol{M}$ according to equation (\ref{eq:hamiltonian}), and $N$ is the number of the nodes, $\boldsymbol{E}$ is an $N\times N$ identity matrix, and $\boldsymbol{M}$ is an $N\times N$ matrix, which all elements are equal 1. One can prove, that the the matrix powers of this $\bs{M}$ matrix can be written in the form $\bs{M}^k=N^{k-1} \bs{M}$ if $k>0$. We can represent the unitary time evolution operator by the form 
\begin{equation}
	\bs{U}(t) = \e^{-i\bs{H}t} = \e^{-iN\bs{E}t} \e^{i\bs{M}t} \, 
\end{equation}
because the identity matrix commutes with $\bs{M}$. The first exponential is trivial, and the second one can be evaluated as
\begin{equation}
	 \makebox[100mm] {$\e^{i\bs{M}t} = \sum_{n=0}^{\infty} \frac{\left ( i \bs{M} t\right )^n}{n!} = 
	\bs{E}+\frac{1}{N} \sum_{n=0}^{\infty} \frac{\left ( i t\right )^n N^n }{n!} \bs{M} - \frac{1}{N} \bs{M} =
	\bs{E} + \frac{1}{N} \e^{iNt} \bs{M} - \frac{1}{N} \bs{M} \, .$}
\end{equation}
Using this equation the time evolution can be described as
\begin{equation}
	\bs{U}(t) = \frac{1}{N} \bigg ( N \e^{-iNt}\bs{E} + \left ( 1-\e^{-iNt} \right )\bs{M} \bigg ) \, .
\end{equation}
The return probability with the above initial state reads
\begin{equation}
	P_1(t) = \frac{(N-1)^2}{N^2} + \frac{1}{N^2} + 2 \frac{N-1}{N^2}\cos(Nt) \, ,
\label{quantP1}
\end{equation}
here we note that this result for the exact return probability is accordance with the average return probability in \cite{Anishchenko}. We also plot this function with rescaled time corresponding to equation (\ref{eq:one}) in figure \ref{fig:complete} with (black) line, we find an excellent match with the numerical simulation.

The introduced simple treatment can be straightforwardly applied for a classical system, corresponding to equation (\ref{classp0}). The return probability in a classical system with the previous conditions reads
\begin{equation}
	P_{1}^{c}(t) = \frac{1}{N} \bigg ( \left ( N-1 \right ) \e^{-Nt} +1  \bigg ) \, .
\label{classP1}
\end{equation}
We also simulate the classical random walk in this system with the same $\lambda$ probability. Our numerical results ((red) squares), and the scaled analytical line ((black) dashed line) can be seen on figure \ref{fig:complete}. We also get an almost exact match between the simulation and the theoretical line. The classical return probability in equation (\ref{classP1}) decays exponentially in time, while in the quantum case (equation (\ref{quantP1})) we find periodically full revivals. In spite of the strong noise, caused by the rapidly changing underlying graph, the continuous-time quantum walk keeps its nonclassical properties in the $\tau\to 0$ limit. 

\section{Finite step-size behavior}
\label{sec:Finite}

In the previous sections we analytically prove and numerically demonstrated that the fast dynamical percolation causes a time rescaling in the dynamics of the continuous-time quantum walks. In our calculation we use the $\tau\to 0$ limit, and in the numerical simulations we choose a small enough $\tau$ time step-size. Of course, it is only a mathematical limit, in a physical system we cannot get this exact limit. In this section we present some concrete examples in order to show that for every time $T$ there exists a small enough $\tau$ parameter, for which we get a very good estimation for the final state. This can be relevant in experimental realizations.

\subsection{Return probability with finite step-size}

\begin{figure}[!ht]
\centering
        \includegraphics[scale=0.9]{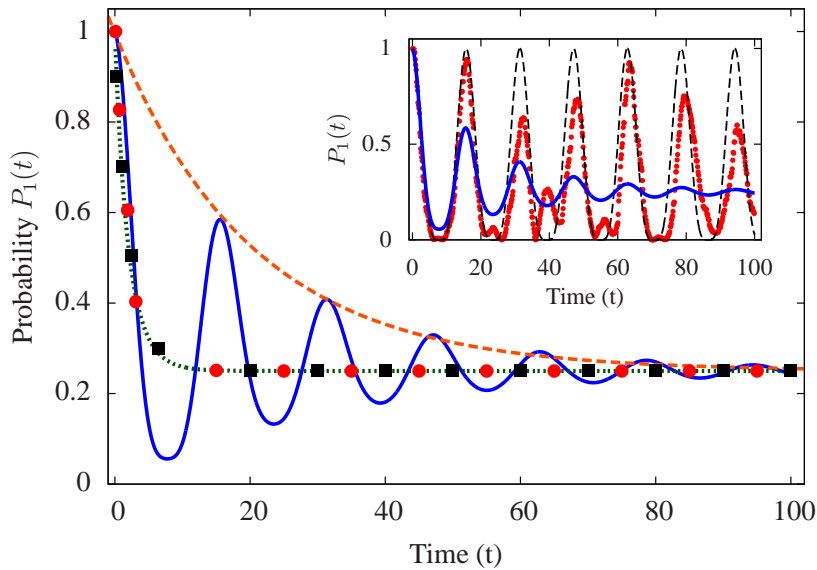}
\caption{The probability of measuring the walker at the starting node on a $N=4$ percolated
	periodical chain with $\lambda=0.2$ percolation probability.
	{\bf Inset}: Dashed (black) line: theoretical function from equation (\ref{eq:supopone}). Points (red): one quantum trajectory with $S=3000$. Solid (blue) line: superoperator formalism with $S=1000$.
	{\bf Main figure:} Solid (blue) line: same as the small figure, and the dotted (orange) line is its envelop. Dashed (green) line: the theoretical $P_1(t)$ probability in a classical system. Circle (red): one classical trajectory with
	$S=1000$. Square (black): superoperator formalism in the classical system with $S=1000$.}
\label{fig:longtime}
\end{figure}

The following simulations show what happens, if we evolve the system for a long time with finite $\tau$ step-size. We take a periodical chain with $N=4$ nodes, and we consider both the quantum and the classical cases. In figure \ref{fig:longtime}.  we compare quantum and classical walks on a percolated graph. We consider the probability that we measure the walker at the starting point if the initial state was $\ket{\Psi(0)}=\ket{1}$.  The percolation probability was chosen $\lambda=0.2$ in each simulation in this figure. The $P_1(t)$ theoretical probability from equation (\ref{eq:one}) reads 
\begin{equation}
    P_1(t) = \cos(\lambda t)^4 \, ,
\label{cosine}
\end{equation}
it is plotted with (black) dashed line. The probability that we measure the walker at the starting state oscillates and can become close to one after arbitrary long time. With (red) dots we plot one quantum trajectory in the percolated graph. We evolve the system with $S=3000$ steps until time $T=100$. After some time from the start the trajectory cannot be well described with the (\ref{eq:one}) formula, each trajectory will follow a different random path. This figure shows one random trajectory, with a new run of the simulation we would get a similar, but different trajectory. We also examined the superoperator formalism for this percolated system, we draw the probability with (blue) solid line. We evolve the system with  $S=1000$ steps. In the superoperator formalism we calculate the state of the system as an average over the possible trajectories. The probability in the long time limit tends to the classical value 
\begin{equation}
    P_1(t\to\infty)=\frac{1}{N} = 0.25
\label{eq:P1t}
\end{equation}
instead of the periodic function in equation (\ref{cosine}) corresponding to the scaled unpercolated case which was introduced in the $\tau\to 0$ limit. The similar asymptotic dynamics has already been considered for discrete-time quantum walks \cite{Kollar}, where non stationary asymptotic states can occur and it seems that similar methods can be applied to the continuous-time walk explaining. In the main figure we plot again the $P_1(t)$ probability with the superoperator formalism (equation \ref{eq:supopone})) in the quantum system with (blue) solid line.

In order to demonstrate the difference between the quantum and the classical case and illustrate the role of the length of the $\tau$ step-size we plot the $P_1(t)$ probability for a classical system ((green) dashed line) and with numerical simulations both a single trajectory ((red) circle) and its averaged value ((black) squares). The analytic form of the probability in the classical case reads 
\begin{equation}
    P_1(t) = 0.25 + \frac{\e^{-2\lambda t}}{2} + \frac{\e^{-4\lambda t}}{4} \, .
\end{equation}
We would like to emphasize that the decay in the classical system is much faster. In the quantum case, we fit an exponential function to the envelop of the average probability. Using the ansatz 
\begin{equation}
	f(t) = a \cdot \e^{-b\cdot t} + 0.25 \, ,
\end{equation}
we find for the fitted parameters $a = 0.746 \pm 0.008$ and $b = 0.049 \pm 0.001$. The  exponent $b$ of the envelope of the decay in the quantum system is almost ten times smaller than the exponent of the classical decay with the same $\tau$ and $\lambda$ parameters. These results demonstrate that there is a significantly different evolution in a quantum system if the $\tau$ step-size of the underlying graph is not small enough. Rescaling the unperturbed evolution is a good approximation only at the beginning of the considered time interval. 

\subsection{Refinement of precision}

\begin{figure}[!ht]
\centering
        \includegraphics[scale=0.9]{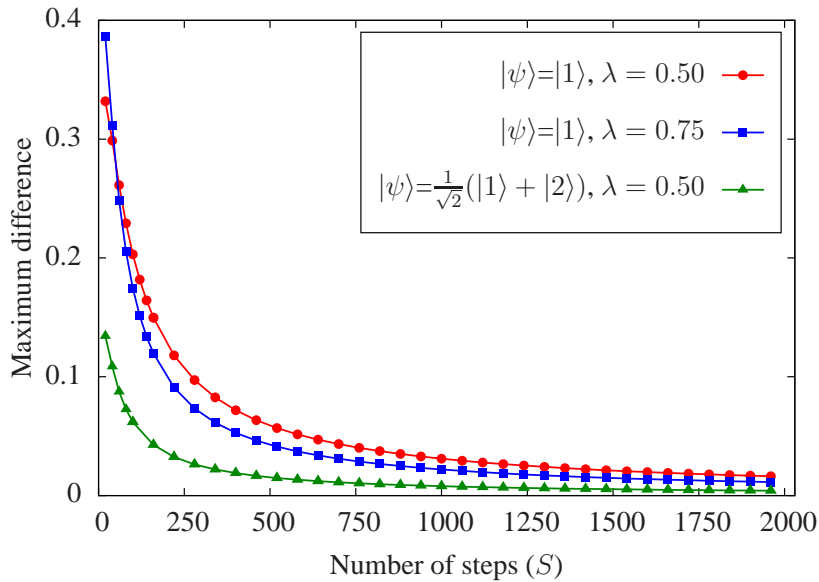}
\caption{The maximum difference between the numerical simulation and the theoretical values from equation (\ref{eq:supopone}) 	of the return probability to the $\ket{1}$ state in an $N=10$ periodical chain for a fixed time interval $T=10$ with 			different initial states and percolation probabilities. Numerical
	values are calculated with the superoperator formalism.}
\label{fig:difference}
\end{figure}

The previous section suggest, that we should be careful with choosing the size of the $\tau$ step-size and the $S$ number of the steps. In the following we quantify the error caused by the finiteness of the step-size.
In figure \ref{fig:difference}. we consider the system for time $T=10$, and we divide the $[0,T]$ interval into $S$ parts, i.e. we evolve the system in $S$ steps with the superoperator formalism until $T$. Therefore if $S$ is big, then the $\tau=T/S$ parameter is small. In our simulation we use a periodical chain with $N=10$ nodes with different initial state and $\lambda$ probabilities. For every $S$ value, we calculate the probability of being at the $\ket{1}$ state both with the numerical simulation and from equation (\ref{eq:supopone}), and we take the maximum difference between them. It is nicely seen that the difference between the simulation and the theory is a decreasing function of the number of the steps. 
\begin{figure}[!ht]
\centering
        \includegraphics[scale=0.9]{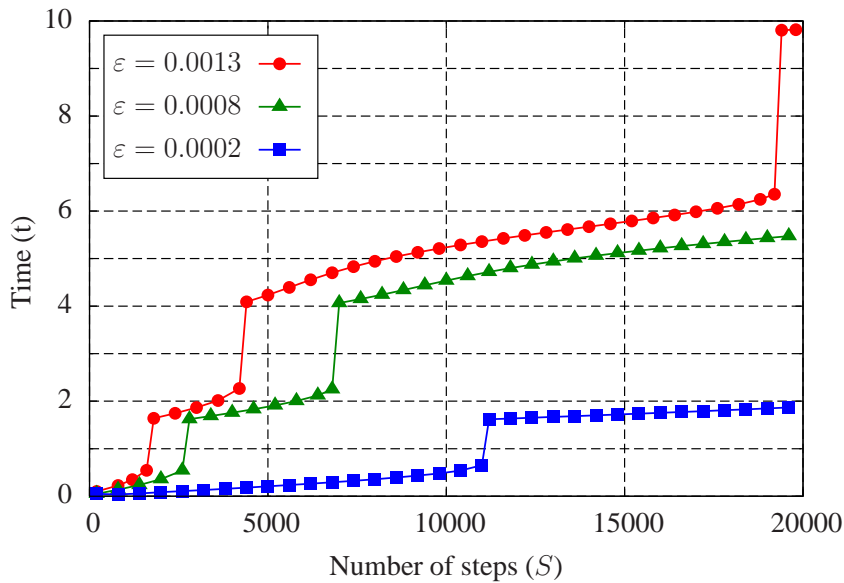}
        \caption{The maximum time while the relative error of the theoretical formula in equation (\ref{eq:supopone}) and the simulated $P_1(t)$ values are smaller than an $\varepsilon$ parameter. We use an $N=5$ periodical chain with the superoperator formalism, and we split the $[0,10]$ interval into $S$ parts, i.e. the step-size was $\tau=10/S$.}
\label{fig:epsilon}
\end{figure}

In our last simulation (figure \ref{fig:epsilon}.) we illustrate, how long one can reliably describe the system via the rescaled evolution with relative error smaller than a given $\varepsilon >0 $ parameter, as a function of the number of time steps. We use a periodical chain with $N=5$ nodes, and we choose the percolation probability for $\lambda=0.5$. We evolve the system with the superoperator formalism from the $\ket{\Psi(0)}=\ket{1}$ state. We take the $[0,T]$, $T=10$ interval, and divide it into $S$ parts, therefore the step-size was $\tau=T/S$. In each step we calculate the relative error of the $P_1(t)$ probability both with the (\ref{eq:supopone}) formula and from the simulation. If the relative difference of these two values becomes bigger or equal than an $\varepsilon$ parameter we stop the simulation. The jumps in the function can be explained by the oscillatory behavior of the $P_1(t)$ function. This figure shows that for every $T$ time and $\varepsilon > 0$ parameter there exist a $\tau$ step-size, for which the relative error of the (\ref{eq:supopone}) (or the (\ref{eq:one})) formula is smaller than $\varepsilon$. It means that our results can be extended for arbitrary large time interval, provided we choose a small enough step-size.

\section{Conclusion}
\label{sec:Conclusion}

In this article we studied the time evolution of the continuous-time quantum walk on a percolation graph. Both unique trajectories and their average properties by using a superoperator formalism have been examined. Strong percolation can be achieved in two limits, leading to different behavior: one can either fix the step-size and take the long time limit, or fix the evolution time and tend to zero with the $\tau$ time scale of changing the graph. We note that the latter limit is not present in the case of discrete-time quantum walks, where the discrete time-step sets a minimum for the step-size of percolation.

In the fast percolation limit the time evolution will be described by a simple analytical formula. In fact, the percolation causes only a time rescaling compared to the percolation free case. This result is independent of  the underlying graph and holds exactly in the $\tau \to 0$ limit. In the other case, a small, but finite $\tau$ is kept fixed and one is interested in the long time limit. In this case, there is an initial period of time when the rescaled, unpercolated evolution is a good approximation, but then the typical oscillations for the probability of finding the particle at a vertex are damped and our numerical simulations suggest that asymptotically one always reaches a flat distribution on a finite graph, unlike in the discrete-time case where the coin state of the walker might oscillate even in the asymptotic limit. We have numerically evaluated the error in the approximation in some examples. The discussed system may serve as a simple model for transport in disordered media.

\section*{Acknowledgments}
        We acknowledge support by the Hungarian Scientific Research Fund (OTKA) under Contract No. K83858,
        the M\"OB-DAAD project No. 40018 and the Hungarian Academy of Sciences (Lend\"ulet Program, LP2011-016).
        We thank Prof. Oliver M\"ulken and Anastasiia Anishchenko for the kind hospitality during a visit  in Freiburg
        and for stimulating discussions.

\section*{References}
\bibliography{refs}
\bibliographystyle{unsrt}

\end{document}